\documentclass[twocolumn]{pasj01}
\Received{$\langle$reception date$\rangle$}
\Accepted{$\langle$acception date$\rangle$}
\Published{$\langle$publication date$\rangle$}
\usepackage{mathpazo}
\usepackage[T1]{fontenc}

\begin{document}

\title{Clump-scale chemistry in the NGC\,2264-D cluster-forming region}
\author{Kotomi \textsc{Taniguchi},\altaffilmark{1,}$^{*}$ Adele \textsc{Plunkett},\altaffilmark{2} Tomomi \textsc{Shimoikura},\altaffilmark{3} Kazuhito \textsc{Dobashi},\altaffilmark{4} Masao \textsc{Saito},\altaffilmark{1,5}  Fumitaka \textsc{Nakamura},\altaffilmark{1,5,6} and Eric \textsc{Herbst}\altaffilmark{7,8}}

\altaffiltext{1}{National Astronomical Observatory of Japan (NAOJ), Osawa, Mitaka, Tokyo 181-8588, Japan}
\altaffiltext{2}{National Radio Astronomy Observatory, 520 Edgemont Rd., Charlottesville, VA 22903, USA}
\altaffiltext{3}{Faculty of Social Information Studies, Otsuma Women's University, Sanban-cho, Chiyoda, Tokyo 102-8357, Japan}
\altaffiltext{4}{Department of Astronomy and Earth Sciences, Tokyo Gakugei University, Nukuikitamachi, Koganei, Tokyo 184-8501, Japan}
\altaffiltext{5}{Department of Astronomical Science, School of Physical Science, SOKENDAI (The Graduate University for Advanced Studies), Osawa, Mitaka, Tokyo 181-8588, Japan}
\altaffiltext{6}{Department of Astronomy, Graduate School of Science, The University of Tokyo, Hongo, Bunkyo, Tokyo 113-0033, Japan}
\altaffiltext{7}{Department of Astronomy, University of Virginia, Charlottesville, VA 22904, USA}
\altaffiltext{8}{Department of Chemistry, University of Virginia, Charlottesville, VA 22904, USA}
\email{kotomi.taniguchi@nao.ac.jp}

\KeyWords{astrochemistry --- ISM: molecules -- stars: formation}

\maketitle

\begin{abstract}
We have conducted mapping observations toward the n3 and n5 positions in the NGC\,2264-D cluster-forming region with the Atacama Compact Array (ACA) of the Atacama Large Millimeter/submillimeter Array (ALMA) in Band 3.
Observations with 10000 au scale beam reveal the chemical composition at the clump scale. 
The spatial distributions of the observed low upper-state-energy lines of CH$_{3}$OH are similar to those of CS and SO, and the HC$_{3}$N emission seems to be predominantly associated with clumps containing young stellar objects.
The turbulent gas induced by the star formation activities produces large-scale shock regions in NGC\,2264-D, which are traced by the CH$_{3}$OH, CS and SO emissions.
We derive the HC$_{3}$N, CH$_{3}$CN, and CH$_{3}$CHO abundances with respect to CH$_{3}$OH.
Compared to the n5 field, the n3 field is farther (in projected apparent distance) from the neighboring NGC\,2264-C, yet the chemical composition in the n3 field tends to be similar to that of the protostellar candidate CMM3 in NGC\,2264-C.
The HC$_{3}$N/CH$_{3}$OH ratios in the n3 field are higher than those in the n5 field.
We find an anti-correlation between the HC$_{3}$N/CH$_{3}$OH ratio and their excitation temperatures.
The low HC$_{3}$N/CH$_{3}$OH abundance ratio at the n5 field implies that the n5 field is an environment with more active star formation compared with the n3 field.
\end{abstract}

\section{Introduction} \label{sec:int}

Clusters are major sites of star formation, and almost 90\% of stars are born in clusters \citep{2003ARA&A..41...57L}.
Hence, it is important to study physical conditions and evolution of clusters for the understanding of star formation.
Furthermore, clusters are essential from the point of view of the formation of the solar system, because several studies suggest that the Sun was born in a cluster region \citep{2010ARA&A..48...47A}.
\citet{2019A&A...631A..25J} compared the HDO/H$_{2}$O ratios among comets in the solar system, clustered Class 0 protostars and isolated Class 0 protostars.
Their results support that the Sun formed in a clustered and relatively warm environment.

The gas-phase chemical composition is known as a good diagnostic tool to investigate the physical conditions and evolutionary stages from starless stage to protostellar stages \citep{2012A&ARv..20...56C, 2020ARA&A..58..727J, 2017ApJS..228...12T, 2019ApJ...872..154T}.
There are largely two types of molecules in star-forming regions: unsaturated carbon-chain species and saturated complex organic molecules (COMs).
Carbon-chain molecules are formed in the gas phase at an early stage of starless cores, and it was classically considered that they are deficient in evolved star-forming cores (e.g., \cite{1992ApJ...392..551S}).
However, recent observations detected cyanoacetylene (HC$_{3}$N), one of the unsaturated carbon-chain species, around many low-mass and high-mass young stellar objects (YSOs) with single-dish survey observations and interferometric observations \citep{2017ApJ...841..120B, 2018ApJ...863...88L, 2018ApJ...854..133T,  2019ApJ...872..154T, 2020ApJ...898...54T}.
On the other hand, COMs are mainly formed on dust surfaces and abundant in hot ($>100$ K) and dense ($\geq 10^{6}$ cm$^{-3}$) gas around YSOs, namely hot core and hot corino chemistry around high-mass and low-mass stars, respectively \citep{2004ASPC..323..195C, 2009ARA&A..47..427H}.
In addition to these two types of molecules, molecules tracing gas kinematics (CO and its isotopologues) and shock tracers (e.g., SiO) are useful indicators for investigation of physical conditions in star-forming regions.

Since carbon-chain molecules and COMs have different dominant formation mechanisms, their abundance ratios may be useful tracers to investigate physical conditions in star-forming regions. 
For example, \citet{2021ApJ...910..141T} derived the CH$_{3}$OH/HC$_{3}$N abundance ratio around three low-mass YSOs and compared them to chemical network simulations to constrain ages of YSOs.
They found that the CH$_{3}$OH/HC$_{3}$N abundance ratio depends on not only the evolutionary stages of YSOs but also physical parameters, especially bolometric luminosity.
These results suggest that the CH$_{3}$OH/HC$_{3}$N abundance ratio is a tracer to investigate environments, e.g., degree of UV radiation, where YSOs are born.

However, chemical processes and the utility of the chemical composition at the larger clump scale or the cluster-forming clump scale is less constrained. 
Some of the gas and dust composing cluster-forming clumps will be involved in the protostellar system, and the initial chemical composition of cores should influence the eventual protostar formation.
Thus, chemical observations at the large cluster-forming-clump scale are needed.
 
\citet{2018ApJ...855...45S} conducted observations of multi-molecular emission lines toward 24 massive cluster-forming clumps with the Nobeyama 45 m radio telescope.
They categorized the observed clumps into four types; clumps without clusters (Type 1), clumps showing good correlations with clusters (Type 2), clumps showing poor correlations with clusters (Type 3), and clusters without associated clumps (Type 4).
It was proposed that clumps evolve from Type 1 through Types 2 and 3, to Type 4, based on analyses of the gas kinematics and the spatial coincidence of gas and star density.
They also investigated relationships between molecular abundance ratios, HC$_{3}$N vs. CCS and SO vs. CS, and the previously mentioned clump evolution.
If the chemical evolution is assumed to progress in the same way as that of the core scale proposed by \citet{1992ApJ...392..551S}, their results imply that Type 1 is chemically more evolved compared to Type 2.
This conclusion derived from the chemical composition is opposite to their clump formation scenario; Type 2 is dynamically more evolved.
Hence, the clump-scale chemistry seems to be different from that of the core scale, and impacts induced from several stars may change chemical compositions in cluster-forming clumps.

NGC\,2264, composed of several cluster-forming clumps, is the second closest high-mass star-forming region ($719^{+22}_{-21}$ pc; \cite{2019A&A...630A.119M}), and a good target source for investigating chemical composition.
The NGC\,2264-D cluster-forming clump is located just north of the NGC\,2264-C cluster-forming clump.
A B2-type object (IRAS 06384+0932), whose bolometric luminosity is $\sim 2300$ L$_{\odot}$, is located in NGC\,2264-C \citep{2006A&A...445..979P}.
Another Class I YSO (IRAS 06382+0939) with a bolometric luminosity of $\sim 150$ L$_{\odot}$ is located in the NGC\,2264-D cluster-forming clump \citep{2006A&A...445..979P}.
These sources are called IRS1 and IRS2, respectively.

Chemical characteristics of the NGC\,2264-C cluster-forming clump have been well studied \citep{2015ApJ...809..162W,2017ApJ...847..108W, 2016MNRAS.458.1742C}.
\citet{2015ApJ...809..162W} conducted 0.8 mm, 3 mm, and 4 mm line survey observations toward CMM3, a high-mass protostellar candidate located in the center of the NGC\,2264-C cluster-forming clump, with the Nobeyama 45 m and ASTE telescopes.
They detected various molecules including COMs, carbon-chain molecules, and deuterated species. 
These characteristics imply that CMM3 in NGC\,2264-C is chemically younger than Orion KL \citep{2015ApJ...809..162W}.
\citet{2017ApJ...847..108W} presented the 0.8 mm line survey data toward CMM3 obtained with the Atacama Large Millimeter/submillimeter Array (ALMA).
These data with an angular resolution of 0\farcs9 spatially resolve CMM3, revealing that it is a binary system candidate.
These two binary cores may have different chemical characteristics; CMM3A is rich in molecular emission, while molecular emission is deficient at CMM3B.
\citet{2016ApJ...822...85L} showed the presence of a complex network of protostellar outflows in NGC\,2264-C using the SiO emission line. 

Although the chemical composition in NGC\,2264-C has been investigated with both single-dish telescopes and ALMA, chemical compositions in NGC\,2264-D have not been studied in detail so far.
Recently, \citet{2020MNRAS.493.2395T} conducted mapping and position-switching observations with the Nobeyama 45 m radio telescope towards NGC\,2264-C and NGC\,2264-D.
The northern edge of NGC\,2264-D appears to be chemically younger than its southern part.
They suggested that the chemical evolution in these regions seems to reflect the combination of evolution along the filamentary structure and evolution of each clump. 

In this paper, we present mapping observations towards two positions in the NGC\,2264-D cluster-forming region conducted by Atacama Compact Array (ACA).
The two positions, n3 and n5, were identified based on HC$_{3}$N emission by \citet{2020MNRAS.493.2395T} (see Fig.~\ref{fig:nobeyama}).
The n3 position is located at the northern edge of NGC\,2264-D and the integrated intensity ratio of HC$_{3}$N and CH$_{3}$OH, $I$(HC$_{3}$N)/$I$(CH$_{3}$OH), at this position shows the highest value among the other positions in NGC\,2264-C and NGC\,2264-D.
A variety in the $I$(HC$_{3}$N)/$I$(CH$_{3}$OH) ratio implies a chemical differentiation in the cluster-forming region, suggestive of different evolutionary histories and/or different environments \citep{2016A&A...592L..11S, 2019ApJ...881...57T, 2020A&A...643A..60S}.
The n3 position is located to the west of the IRS2 source (IRAS 06382+0939).
The n5 position lies at the southern part of the NGC\,2264-D cluster-forming clump and shows the highest column density ratios of N$_{2}$H$^{+}$/CCS and N$_{2}$H$^{+}$/HC$_{3}$N among the other positions in NGC\,2264-D, which means that the n5 position is the most chemically evolved position in this cluster-forming clump.
In Section~\ref{sec:obs}, the observations and reduction procedures are described.
The dust continuum ($\lambda=3$ mm) and moment 0 maps of molecular lines are presented in Section~\ref{sec:map}, and spectra are shown in Section~\ref{sec:spec}, respectively.
We analyze spectra using the CASSIS software \citep{2015sf2a.conf..313V} in Section~\ref{sec:ana}.
We compare spatial distributions of HC$_{3}$N and CH$_{3}$OH, and chemical compositions of the n3 field, the n5 field, and CMM3 in NGC\,2264-C in Sections~\ref{sec:d1} and ~\ref{sec:d2}, respectively.
Main conclusions of this paper are summarized in Section~\ref{sec:con}.

\section{Observations and Data Reduction} \label{sec:obs}

\begin{figure}
       \begin{center}
	\includegraphics[scale=0.65, bb = 0 20 305 400]{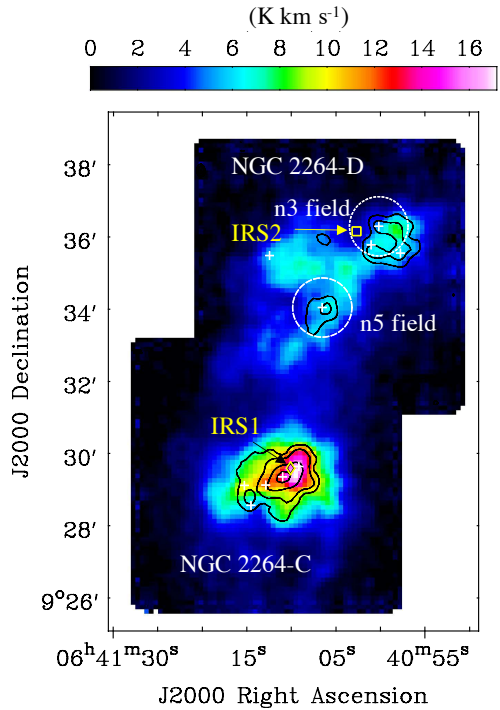}
	\end{center}
       \caption{Moment 0 maps of HNC ($J=1-0$) and HC$_{3}$N ($J=10-9$) lines, as shown in the color scale and black contours, respectively, toward the NGC\,2264-C and NGC\,2264-D cluster-forming clumps obtained with the Nobeyama 45 m telescope \citep{2020MNRAS.493.2395T}. Black contour levels are 15\%, 25\%, 50\%, 75\%, and 90\% of the peak levels (6.65 K \,km\,s$^{-1}$). The white dashed circles represent the FoV of the ALMA ACA data (97\arcsec) presented in this paper. The white crosses indicate positions analyzed by \citet{2020MNRAS.493.2395T}.}
    \label{fig:nobeyama}
\end{figure}

We analyzed the ALMA ACA Cycle 7 data in Band 3 toward the n3 and n5 positions in the NGC\,2264-D cluster-forming clump\footnote{Project ID; 2019.2.00030.S, PI; Kotomi Taniguchi}.
Fig.~\ref{fig:nobeyama} shows the moment 0 map of HNC (color scale) overlaid by the HC$_{3}$N moment 0 map (black contours) toward the NGC 2264-C and NGC 2264-D cluster-forming clumps, as obtained with the Nobeyama 45 m radio telescope \citep{2020MNRAS.493.2395T}.
The white dashed circles indicate fields observed by the ALMA ACA observations.

The ALMA ACA observations were carried out between January 1st and January 16th in 2020.
The phase reference centers were ($\alpha_{2000}$, $\delta_{2000}$) = (6$^{\rm h}$41$^{\rm m}$00\fs16, +9\arcdeg36\arcmin18\farcs0) and (6$^{\rm h}$41$^{\rm m}$06\fs52, +9\arcdeg34\arcmin03\farcs1) for the n3 and n5 positions, respectively.
Hereafter, we call each field of view the ``n3 field'' and ``n5 field'', respectively.
The phase calibrator source was J0643+0857 during all of the observations.
For bandpass and flux calibrations, J0725-0054 and J0854+2006 were observed during the n3 observations, and only J0725-0054 was observed during the n5 observations.
The Field of View (FoV) and maximum recoverable scale (MRS) are $\sim$97\arcsec and $\sim$65\arcsec, respectively.

%%%%%%%%
%%%%%%%%%%

Table~\ref{tab:SpW} summarizes information for each spectral window.
The bandwidth and frequency resolution for the continuum observation are 1875 MHz and 1.129 MHz, respectively.
The bandwidths are 58.59 MHz and 117.19 MHz for the spectral window including the CH$_{3}$OH lines, and the spectral windows including the other lines (HC$_{3}$N and CH$_{3}$CN), respectively.
The frequency resolution for these molecular lines is $\sim 141$ kHz, corresponding to a velocity resolution of $\sim 0.38-0.44$ \,km\,s$^{-1}$.

%%%%%%%%%%%%%%%%%
\begin{table*}
	%\centering
	\caption{Summary of spectral windows}
	\label{tab:SpW}
	\begin{tabular}{llccc} 
		\hline
		Species & Transition & Frequency  & Frequency  \\
		             &                & range (GHz) & resolution (MHz) \\
		\hline
		Continuum$^{a}$ & & $97.6068-99.5591$ & 1.129 \\
		CH$_{3}$OH & $2_{-1,2}-1_{-1,1}$ E & $96.7088-96.7697$ & 0.141 \\
		 & $2_{0,2}-1_{0,1}$ A$^{+}$ & & \\
		HC$_{3}$N & $J=12-11$ & $109.097-109.219$ & 0.141 \\
		CH$_{3}$CN & $J=6-5$, $K=0,1$ & $110.300-110.422$ & 0.141 \\
		\hline
		\multicolumn{4}{l}{$a$ This spectral window contains the CS, SO, and CH$_{3}$CHO lines.} \\
	\end{tabular}
\end{table*}
%%%%%%%%%%%%%%%%%

We conducted data reduction and imaging using the Common Astronomy Software Application (CASA v 5.6.1; \cite{2007ASPC..376..127M}) on the pipeline-calibrated visibilities.
The data cubes were made with the CASA tclean task.
Uniform weighting was applied.
The resulting angular resolution is approximately 17\arcsec $\times$ 11\arcsec, corresponding to $\sim$ 0.059 pc $\times$ 0.038 pc ($\sim$ 12200 au $\times$ 7900 au) at the source distance (719 pc; \cite{2019A&A...630A.119M}).
The pixel size and image size are 2\farcs5 and 250 $\times$ 250 pixels.
Continuum ($\lambda=3$ mm) images were created from the data cubes using the IMCONTSUB task.
The polynomial order of 0 was adopted for the continuum estimation.
The noise levels of the continuum images are 1.0\,mJy\,beam$^{-1}$ and 0.9\,mJy\,beam$^{-1}$ at the n3 and n5 fields, respectively.

\section{Results and Analyses} \label{sec:res}

\subsection{Continuum and Moment 0 Maps} \label{sec:map}

%%%%%%%%
\begin{figure*}
       \begin{center}
	\includegraphics[scale=0.65, bb = 0 10 578 300]{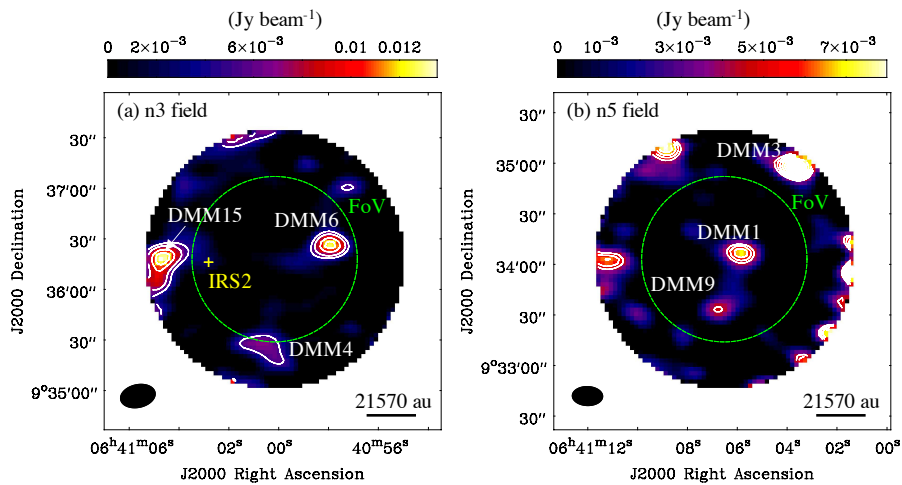}
	\end{center}
       \caption{Continuum ($\lambda=3$ mm) images with a primary beam correction towards (a) the n3 field and (b) the n5 field in the NGC\,2264-D cluster-forming clump. The color scales are adjusted to the maximum values for the peak of clumps in FoV. The contour levels are 5, 7, 10, 12 $\sigma$ for panel (a) and 5, 6, 7, 8, 9, 10 $\sigma$ for panel (b), where the noise levels are around 1.0 and 0.9 \,mJy\,beam$^{-1}$ for panels (a) and (b), respectively. %of the peak levels of the target cores, with peak intensities of 12.3 and 5.43 \,mJy\,beam$^{-1}$ for panels (a) and (b), respectively. 
       %The noise levels are around 1.0 and 0.9 \,mJy\,beam$^{-1}$ for panels (a) and (b), respectively. 
       The green dashed circles represent FoV sizes (97\arcsec). The yellow cross indicates the IRS2 (IRAS 06382+0939) position. The indications of ``DMMn'', where n is a number, denote names of continuum cores identified by \citet{2006A&A...445..979P}. The filled black ellipses indicate angular resolutions of approximately 16\farcs6 $\times$ 10\farcs4 and 17\farcs6 $\times$ 10\farcs2 for panels (a) and (b), respectively. The linear scale is given for 30\arcsec corresponding to 21570 au.}
    \label{fig:cont}
\end{figure*}
%%%%%%%%%%

Fig.~\ref{fig:cont} shows the continuum ($\lambda=3$ mm) images with a primary beam correction toward the two positions in NGC\,2264-D.
Panels (a) and (b) of Fig.~\ref{fig:cont} indicate the n3 and n5 fields, respectively.
One prominent continuum core is located at the position northwest of the field of view in the n3 field.
This core corresponds to DMM6 as identified by \citet{2006A&A...445..979P} using the 1.2 mm dust continuum data.
Additionally, the DMM4 and DMM15 cores have been detected in the n3 field. %at the south and east positions, respectively.
No continuum core was detected at the position of IRS2 (Class I YSO), which is consistent with the 1.2 mm dust continuum data.

In the n5 field, we adjust the color scale to the peak value for the DMM1 core.
Two cores, DMM1 and DMM9 \citep{2006A&A...445..979P}, have been detected near the phase reference center.
Since only a part of the brightest core (DMM3) is detected at the edge of the field, we do not include the DMM3 core in the following discussion.

%%%%%%%%
\begin{figure*}
       \begin{center}
	\includegraphics[scale=0.55, bb=0 10 715 970]{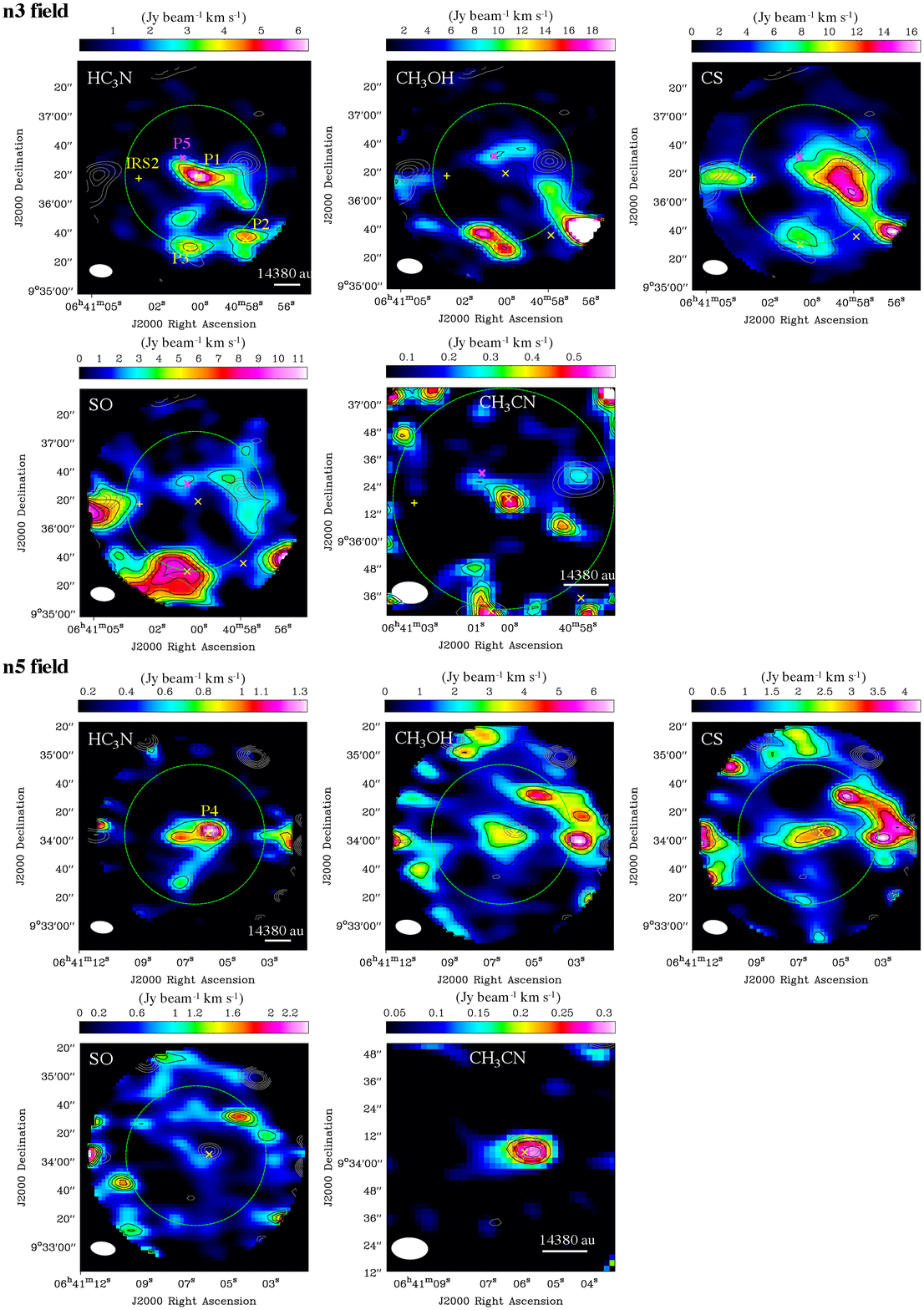}
	\end{center}
       \caption{Moment 0 maps of HC$_{3}$N, CH$_{3}$OH, CS, SO, and CH$_{3}$CN toward the n3 (the first and second raws) and n5 (the third and fourth raws) fields in NGC\,2264-D. The black contour levels are 20\%, 40\%, 60\%, and 80\% of the peak intensities summarized in Table~\ref{tab:mom0} for panels of HC$_{3}$N and CH$_{3}$OH. For panels of CS, SO and CH$_{3}$CN, the black contour levels are [10, 15, 20, 25, 30$\sigma$], [5, 7, 10, 13, 15, 17$\sigma$], [5, 6, 7, 8, 9 $\sigma$] in the n3 field, [5, 7, 9, 11, 13 $\sigma$], [5, 6, 7, 8$\sigma$], and [5, 6, 7$\sigma$] in the n5 field, respectively. The rms noise levels for each panel are summarized in Table~\ref{tab:mom0}. The integrated velocity ranges are $+2-+9$\,km\,s$^{-1}$ for HC$_{3}$N, $-10- +18$\,km\,s$^{-1}$ for CH$_{3}$OH, $-10-+20$\,km\,s$^{-1}$ for CS and SO, and $0-+14$\,km\,s$^{-1}$ for CH$_{3}$CN maps, respectively. The gray contours show the distribution of dust continuum emission, as in Fig.~\ref{fig:cont}. The filled white ellipses represent the angular resolution ($\sim$ 17\arcsec $\times$ 11\arcsec). The yellow crosses indicate the IRS 2 and HC$_{3}$N peak (P1 -- P4) positions, and the magenta cross in the n3 field indicates the CH$_{3}$OH Peak (P5). The green dashed circle indicate the FoV. The linear scales are given for 20\arcsec corresponding to 14380 au.} 
%CS ($J=2-1$; 97.980953 GHz), SO ($3_{2}-2_{1}$; 99.299905 GHz)
    \label{fig:mom0}
\end{figure*}
%%%%%%%%%%
%%%%%%%%
\begin{figure}
       \begin{center}
	\includegraphics[scale=0.4]{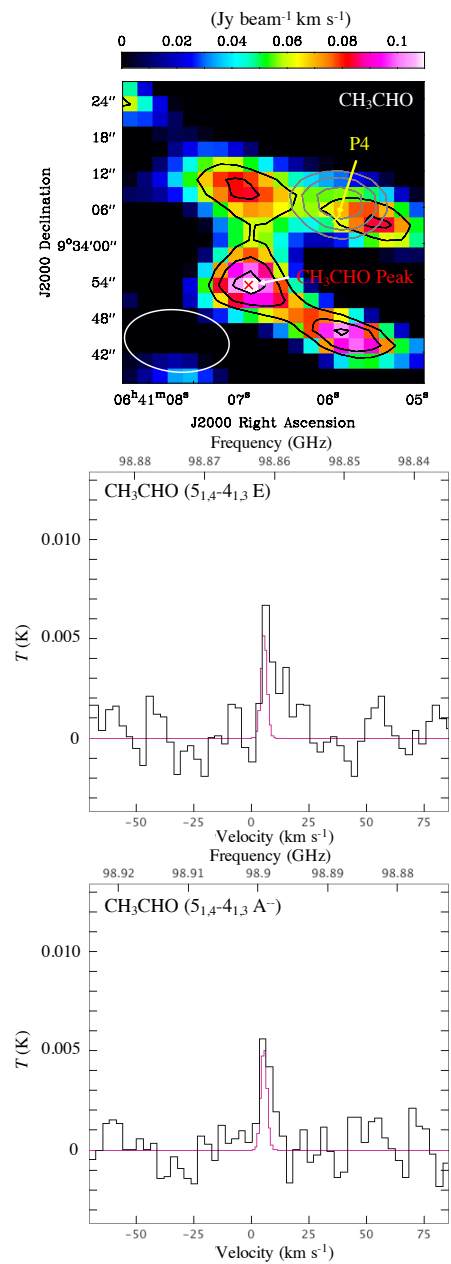}
	\end{center}
       \caption{Top panel: Moment 0 map of CH$_{3}$CHO ($5_{1,4}-4_{1,3}$ E) toward the n5 field. The contour levels are 3, 4, 5$\sigma$. The noise level is 0.02 \,Jy\,beam$^{-1}$ \,km\,s$^{-1}$. The integrated velocity range is $0 - +24$\,km\,s$^{-1}$. The open white ellipses represent an angular resolution of 17\farcs4 $\times$ 10\farcs3. The P4 position indicates the HC$_{3}$N peak position. Middle and bottom panels: Spectra of the CH$_{3}$CHO lines toward the CH$_{3}$CHO Peak. The spectra are obtained by averaging over pixels within the 10\arcsec diameter (7190 au) centered at the CH$_{3}$CHO Peak. The rms noise level of these spectra is $\sim2$ mK. The purple curves indicate the fitting results with the CASSIS software (Section \ref{sec:ana}).}
    \label{fig:CH3CHO}
\end{figure}
%%%%%%%%%%

%%%%%%%%%%%%%%%%%
\begin{table}
	%\centering
	\caption{Summary of moment 0 maps}
	\label{tab:mom0}
	\begin{tabular}{clcc} 
		\hline
		Field & Species & Peak$^{a}$ & RMS$^{a}$ \\
		\hline
		n3  & HC$_{3}$N & 6.29 & 0.1 \\
		      & CH$_{3}$OH & 18.1 & 1.0 \\
		      & CS & 16.7 & 0.5 \\
		      & SO & 11.5 & 0.5 \\
		      & CH$_{3}$CN & 0.55 & 0.05 \\
		n5  & HC$_{3}$N & 1.34 & 0.05 \\
		      & CH$_{3}$OH & 6.59 & 0.3 \\
		      & CS & 4.3 & 0.3 \\
		      & SO & 2.4 & 0.2 \\
		      & CH$_{3}$CN & 0.3 & 0.04 \\	          
		\hline
		\multicolumn{4}{l}{$a$ Unit is \,Jy\,beam$^{-1}$ \,km\,s$^{-1}$.} \\
	\end{tabular}
\end{table}
%%%%%%%%%%%%%%%%%

Fig.~\ref{fig:mom0} shows moment 0 maps (with a primary beam correction) of molecular emission lines -- HC$_{3}$N ($J=12-11$), CH$_{3}$OH (the integrated frequency range covers three lines; $2_{-1,2}-1_{-1,1}$ E, $2_{0,2}-1_{0,1}$ A$^{+}$, and $2_{0,2}-1_{0,1}$ E, but the last line was detected only near the DMM1 core), CS ($J=2-1$), SO ($N_{J}=3_{2}-2_{1}$), CH$_{3}$CN ($J=6-5$, $K=0,1$) -- towards the n3 (the first and second raws) and n5 (the third and fourth raws) fields, respectively.
The CH$_{3}$CN maps have a different size from the others, because the moment 0 maps of CH$_{3}$CN are noisier than the other maps, especially outside the FoV, and its emission is compact.
We therefore show the close-up maps.
Table~\ref{tab:mom0} summarizes peak intensity and rms noise level for each panel.
The CS and SO lines lie in the spectral window of the widest frequency coverage (Table~\ref{tab:SpW}), and the velocity resolution of these lines is 3.43 \,km\,s$^{-1}$.
Such a low velocity resolution does not resolve the detected lines sufficiently, but we can identify these lines clearly.
In the panel of CH$_{3}$OH toward the n3 field, we adjust the color scale to the maximum value for the peak around the DMM4 continuum core so that its morphology in the field can be clearly seen.  
At the west edge where the CH$_{3}$OH emission is saturated, the CH$_{3}$OH lines show broad line features, suggestive of the shock regions, which may be the source of the high integrated intensity.

We identified three and one HC$_{3}$N clumps in the n3 and n5 fields, respectively, and they are labelled as P1--P4.
In the n3 field, we picked up the three clumps with a peak intensity greater than 4 \,Jy\,beam$^{-1}$ \,km\,s$^{-1}$.
We applied 2D Gaussian fitting to the HC$_{3}$N moment 0 maps.
The fit results are summarized in Table~\ref{tab:2Dgaus}. 
P1, P2, and P3 in the n3 field correspond to n3, n2, and n4 positions, respectively, and P4 corresponds to the n5 position identified by \citet{2020MNRAS.493.2395T}.
The HC$_{3}$N emission is not coincident with the strongest dust continuum core (DMM6) in the n3 field, whereas the HC$_{3}$N emission is associated with the strongest 3 mm continuum core, DMM1, in the n5 field.
P3 is associated with the DMM4 dust continuum core, but P2 does not seem to coincide with the dust continuum core in the n3 field.
In the n5 field, a tail-like feature, which is elongated to the east direction, is associated with the P4 peak, and a weaker peak was seen at the east position.
Another weaker HC$_{3}$N emission is associated with the DMM9 core.
The diffuse HC$_{3}$N emission appears to bridge DMM1 and DMM9.

The CH$_{3}$OH spatial distributions are different from those of HC$_{3}$N in both fields.
In the n3 field, an elongated feature is seen around P3. 
Two CH$_{3}$OH peaks in this structure are not coincident with the P3 HC$_{3}$N peak.
The CH$_{3}$OH emission seems to be spatially anti-correlated with the DMM6 continuum core.
The CH$_{3}$OH emission is weak at the dust continuum peak. 
We determined a CH$_{3}$OH peak by eye and named it P5, which is located near P1, as shown in the panel of the n3 field. 
The coordinate of the P5 position is ($\alpha_{2000}$, $\delta_{2000}$) = ($6^{\rm {h}}41^{\rm {m}}$00\fs78, +9\arcdeg36\arcmin29\farcs88).
The emission near P3 likely originates from the molecular outflow due to strong wing emission (Fig.~\ref{fig:specCH3OH}), and therefore the comparison of chemical composition focuses on the region around P5 instead.
In the n5 field, the CH$_{3}$OH emission is associated with the DMM1 continuum core, but does not coincide exactly with the dust peak position.
In addition, a strong peak is located to the west of the DMM1 core. 

The SO ($3_{2}-2_{1}$; 99.299905 GHz) emission shows an elongated feature in the north-eastern to south-western direction, centering on the DMM6 core.
Such a feature resembles the CH$_{3}$OH emission.
The CS ($J=2-1$; 97.980953 GHz) emission shows strong emission near the DMM6 core. 
Both CS and SO show strong features at the edge of the field near P2, which is consistent with the CH$_{3}$OH emission.
In addition, the CS and SO emission have been detected at P3 and at a position east of IRS2.
In the n5 field, the spatial distributions of CS and SO emission lines are roughly similar to that of CH$_{3}$OH. 

The last panels of Fig.~\ref{fig:mom0} shows moment 0 maps of CH$_{3}$CN.
Its peak positions are consistent with the P1 and P4 HC$_{3}$N peaks in the n3 and n5 fields, respectively.
In the n3 field, another core is found to the west of P1, and weak emission is associated with the DMM6 core.

Two CH$_{3}$CHO lines, $5_{1,4}-4_{1,3}$ E (98.8633135 GHz; $E_{\rm {up}}=16.6$ K) and $5_{1,4}-4_{1,3}$ A$^{--}$ (98.9009445 GHz; $E_{\rm {up}}=16.5$ K), have been detected toward the n5 field in the spectral window with the widest frequency band (Continuum observations; Table~\ref{tab:SpW}).
Fig.~\ref{fig:CH3CHO} shows its moment 0 map ($5_{1,4}-4_{1,3}$ E) and spectra of the two lines toward its peak position, determined by eye and indicated as ``CH$_{3}$CHO Peak'' in the moment 0 map. 
The coordinate of the CH$_{3}$CHO Peak is ($\alpha_{2000}$, $\delta_{2000}$) = ($6^{\rm {h}}41^{\rm {m}}$06\fs821, +9\arcdeg33\arcmin53\farcs05).
The spectra are obtained by averaging over pixels within the 10\arcsec diameter centered at the CH$_{3}$CHO Peak.
The CH$_{3}$CHO emission is detected within the CH$_{3}$OH emitting region.

%%%%%%%%%%%%%%%%%
\begin{table*}
	%\centering
	\caption{Summary of 2D gaussian fitting for the HC$_{3}$N moment 0 maps}
	\label{tab:2Dgaus}
	\begin{tabular}{ccccc} 
		\hline
		Name & Peak position (J2000) & Size$^{a}$ & Position angle & Peak intensity  \\
				    & 				      &        & 				&	(\,Jy\,beam$^{-1}$ \, \,km\,s$^{-1}$) \\
		\hline
		P1 & $6^{\rm {h}}41^{\rm {m}}$00\fs021 $\pm$0\fs039, +9\arcdeg36\arcmin17\farcs92 $\pm$0\farcs25 & 30\farcs59 $\pm$ 1\farcs63 $\times$ 11\farcs78 $\pm$ 0\farcs67 & 69.6\arcdeg $\pm$ 2.0\arcdeg & $6.06 \pm 0.21$ \\
		P2 & $6^{\rm {h}}40^{\rm {m}}$57\fs868 $\pm$0\fs051, +9\arcdeg35\arcmin34\farcs48 $\pm$0\farcs38 & 20\farcs9 $\pm$ 2\farcs4 $\times$ 13\farcs8 $\pm$ 1\farcs5 & 97\arcdeg $\pm$ 14\arcdeg & $4.42 \pm 0.27$ \\
		P3 & $6^{\rm {h}}41^{\rm {m}}$00\fs544 $\pm$0\fs074, +9\arcdeg35\arcmin28\farcs77 $\pm$0\farcs35 & 34\farcs7 $\pm$ 2\farcs9 $\times$ 13\farcs3 $\pm$ 1\farcs1 & 96.8\arcdeg $\pm$ 2.9\arcdeg & $4.13 \pm 0.23$ \\
		P4 & $6^{\rm {h}}41^{\rm {m}}$05\fs895 $\pm$0\fs034, +9\arcdeg34\arcmin04\farcs18 $\pm$0\farcs29 & 23\farcs94 $\pm$ 1\farcs52 $\times$ 17\farcs38 $\pm$ 0\farcs98 & 95.3\arcdeg $\pm$ 9.1\arcdeg & $1.27 \pm 0.05$ \\
		\hline
		\multicolumn{4}{l}{$a$: Reported results are deconvolved sizes, which take into account the synthesized beam.} \\
	\end{tabular}
\end{table*}
%%%%%%%%%%%%%%%%%

\subsection{Spectra} \label{sec:spec}
 
\begin{figure*}
       \begin{center}
	\includegraphics[bb = 0 5 478 720, scale=0.7]{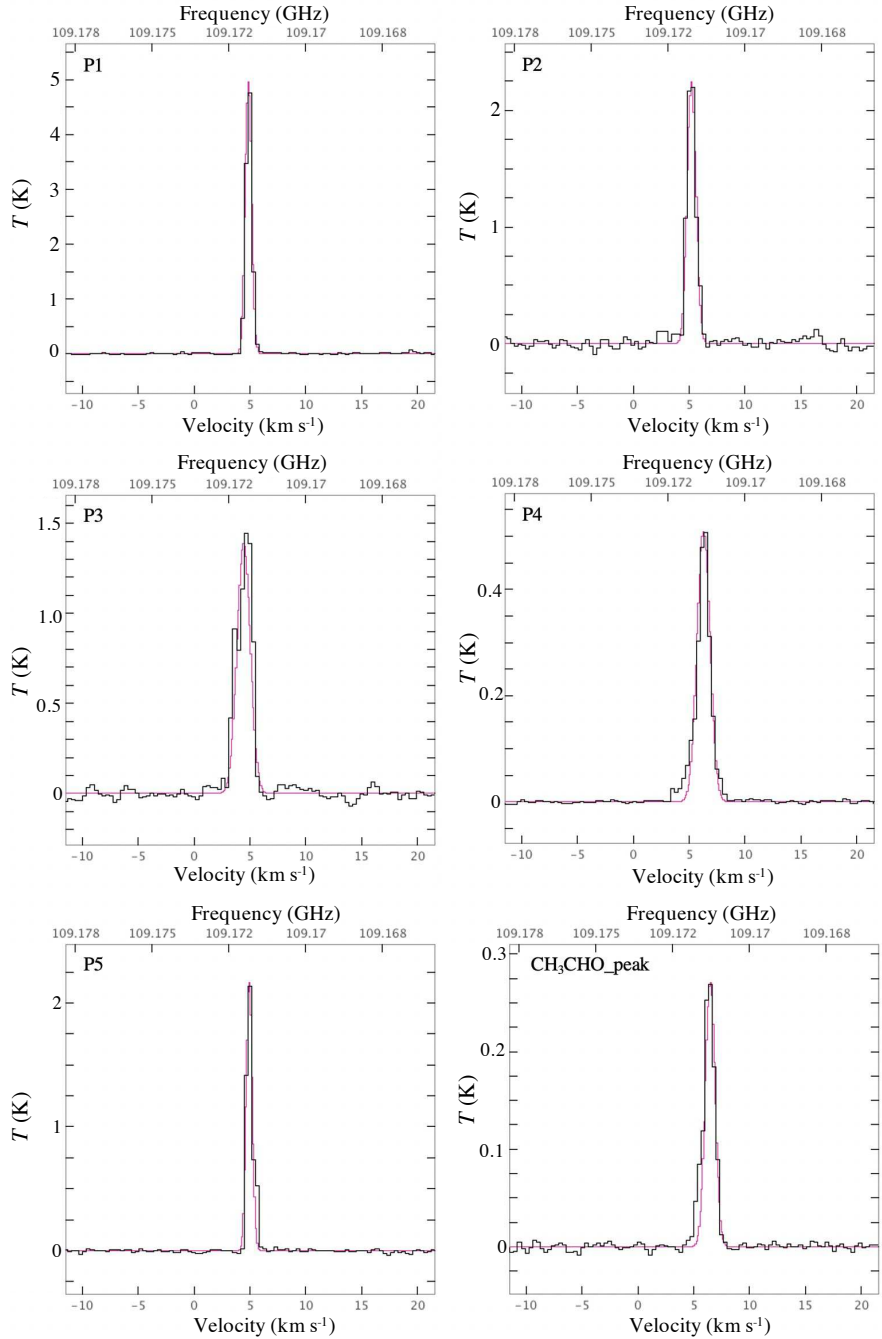}
	\end{center}
       \caption{Spectra of the HC$_{3}$N ($J=12-11$; 109.173634 GHz) line toward the P1--P5 and CH$_{3}$CHO Peak positions. The spectra are obtained by averaging over pixels within the 15\arcsec diameter ($\sim10800$ au) at the centers of P1--P5 peak positions (Table~\ref{tab:2Dgaus}), and 10\arcsec diameter (7190 au) centered at the CH$_{3}$CHO Peak, respectively. The rms noise level of these spectra is $\sim10$ mK and $\sim5$ mK in the n3 field and n5 field, respectively. Purple curves indicate the fitting results using the CASSIS software (Section \ref{sec:ana}).}
    \label{fig:specHC3N}
\end{figure*}

\begin{figure*}
       \begin{center}
	\includegraphics[bb = 0 5 484 720, scale=0.7]{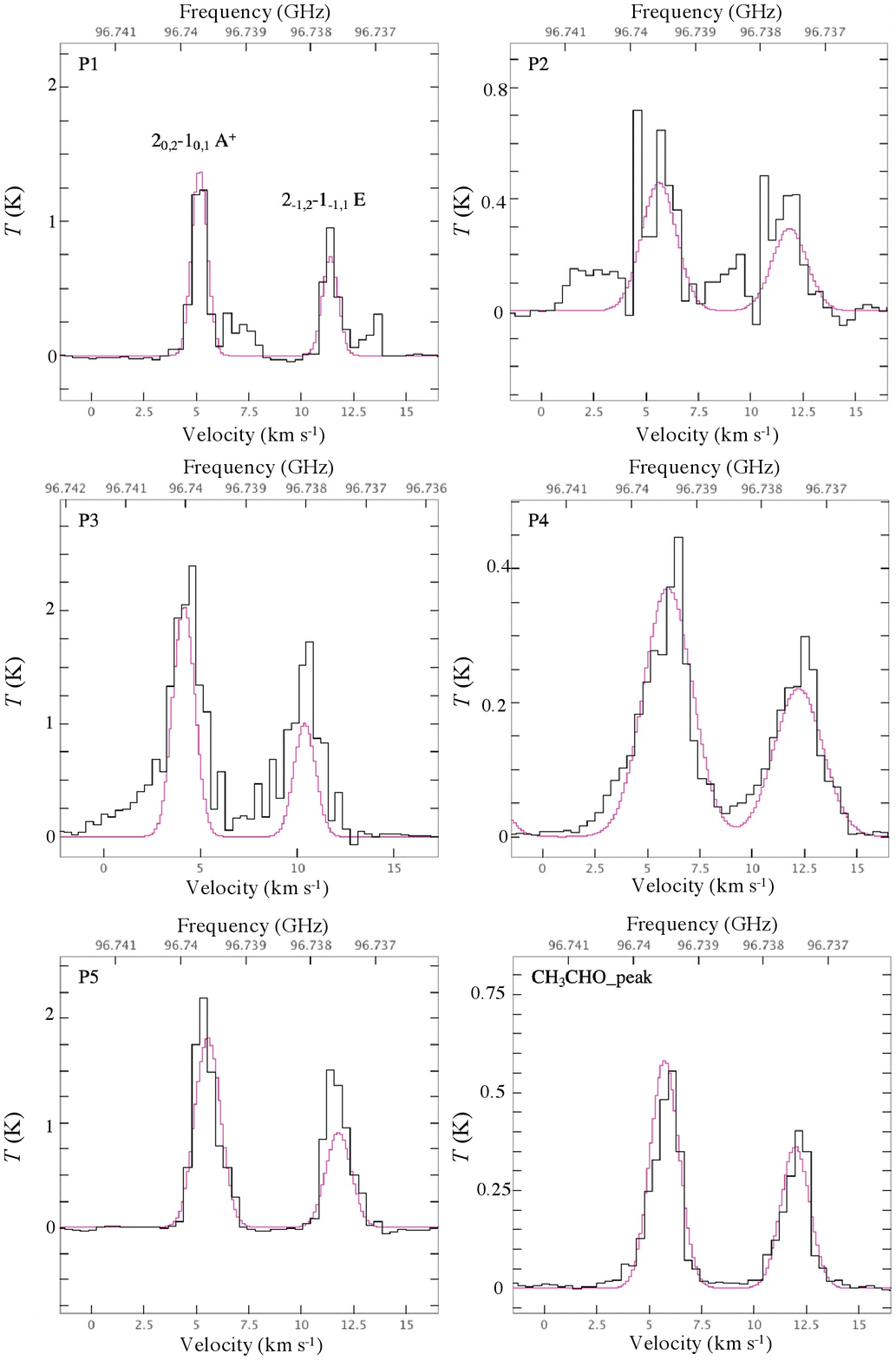}
	\end{center}
       \caption{Spectra of the CH$_{3}$OH ($2_{-1,2}-1_{-1,1}$ E; 96.739358 GHz, and $2_{0,2}-1_{0,1}$ A$^{+}$; 96.741371 GHz) lines toward the P1--P5 and CH$_{3}$CHO Peak positions. The spectra are obtained by averaging over pixels within the 15\arcsec diameter ($\sim10800$ au) at the centers of P1--P5 peak positions (Table~\ref{tab:2Dgaus}), and 10\arcsec diameter (7190 au) centered at the CH$_{3}$CHO Peak, respectively. The rms noise level of these spectra is $\sim9$ mK and $\sim3$ mK in P1 and P4, respectively. Purple curves indicate the fitting results using the CASSIS software (Section \ref{sec:ana}).}
    \label{fig:specCH3OH}
\end{figure*}

\begin{figure*}
       \begin{center}
	\includegraphics[bb = 0 5 540 254, scale=0.7]{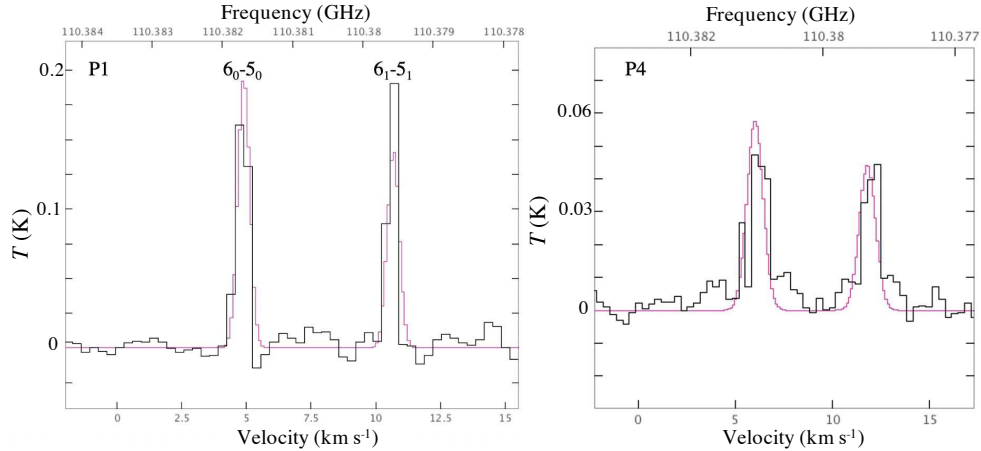}
	\end{center}
       \caption{Spectra of the CH$_{3}$CN ($J_{K} = 6_{0}-5_{0}$; 110.3835 GHz, and $6_{1}-5_{1}$; 110.381372 GHz) lines toward the P1 and P4 positions. The spectra are obtained by averaging over pixels within the 15\arcsec diameter ($\sim10800$ au) at the centers of P1 and P4 peak positions (Table~\ref{tab:2Dgaus}). The rms noise level of these spectra is $\sim9$ mK and $\sim4$ mK in the n3 field and n5 field, respectively. Purple curves indicate the fitting results using the CASSIS software (Section \ref{sec:ana}).}
    \label{fig:specCH3CN}
\end{figure*}

Figs.~\ref{fig:specHC3N} and \ref{fig:specCH3OH} show the spectra of HC$_{3}$N and CH$_{3}$OH, respectively at positions P1--P5, as well as the CH$_{3}$CHO Peak.
The CH$_{3}$CN spectra at P1 and P4 are shown in Fig~\ref{fig:specCH3CN}.
The spectra are obtained by averaging over pixels within the 15\arcsec diameter ($\sim10800$ au) at the centers of P1--P5 peak positions (Table~\ref{tab:2Dgaus}), and 10\arcsec diameter (7190 au) centered at the CH$_{3}$CHO Peak, respectively.
The beam size of 15\arcsec corresponds to the clump size obtained by the 2D Gaussian fitting (Table~\ref{tab:2Dgaus}).

The HC$_{3}$N spectra show narrow line features (Fig.~\ref{fig:specHC3N}). 
A weak wing emission is detected at the P4 position.
The line widths of CH$_{3}$CN are also narrow and comparable with those of HC$_{3}$N.
Since these lines have low upper-state energies ($E_{\rm {up}}$ = 18.5 K and 25.7 K for the $K=0$ and $K=1$ lines, respectively), they seem to originate from outer envelopes, rather than hot core regions close to protostars.

The CH$_{3}$OH spectra show two velocity components at the P1 position, with broad line feature or wing emission, as seen in Fig.~\ref{fig:specCH3OH}.
While the second velocity component may be detected at P2, the current noise level is not low enough to confirm it.
The strong wing emission has been detected at P3.
These results support the premise that the observed CH$_{3}$OH lines trace shock regions.
The missing fluxes were estimated at $\sim10$\% and $\sim30$\% in the n3 and n5 fields, respectively, determined by comparisons to the CH$_{3}$OH data obtained with the Nobeyama 45-m telescope \citep{2020MNRAS.493.2395T}.

%%%%%%%%
%%%%%%%%%%

%%%%%%%%
%%%%%%%%%%

%%%%%%%%
%%%%%%%%%%

\subsection{Spectral analysis with CASSIS} \label{sec:ana}

We analyzed spectra of HC$_{3}$N (Fig.~\ref{fig:specHC3N}), CH$_{3}$OH (Fig.~\ref{fig:specCH3OH}), and CH$_{3}$CN (Fig.~\ref{fig:specCH3CN}) with the CASSIS software \citep{2015sf2a.conf..313V} and the JPL spectroscopic database\footnote{<https://spec.jpl.nasa.gov>}.
Since we have only one line or two lines with similar upper-state energies for each species, we applied the Markov chain Monte Carlo (MCMC) method assuming the local thermodynamic equilibrium (LTE) model available in CASSIS.
In the MCMC method, the minimum and maximum values set for a parameter in a component define the bounds in which the values are chosen randomly. 
The column density ($N$), excitation temperature ($T_{\rm {ex}}$), line width (FWHM), and radial velocity ($V_{\rm {LSR}}$) were derived and treated as semi-free parameters within certain ranges.
CASSIS iteratively proceeds through a given parameter space via a random walk and reaches the final solution by a $\chi^{2}$ minimization\footnote{<http://cassis.irap.omp.eu/docs/CassisScriptingDoc/>}.

Taking previous observational results of $T_{\rm {ex}}$ of N$_{2}$H$^{+}$ with a beam size of 18\arcsec ($\sim 10-26$ K; \cite{2020MNRAS.493.2395T}), we set the range of $T_{\rm {ex}}$ between 10 and 40 K so that this $T_{\rm {ex}}$ range fully covers the range of the N$_{2}$H$^{+}$ excitation temperature in the NGC\,2264-D cluster-forming region. 
The ranges of other parameters are summarized in Table~\ref{tab:a1} in the Appendix.

Table~\ref{tab:CASSIS1} summarizes results derived by the MCMC method.
Since parts of P2 and P3 clumps are located outside the FoV, the derived column densities can be erroneous, and we omit a quantitative discussion in the remainder this paper.
The derived $T_{\rm {ex}}$ values are consistent with $T_{\rm {ex}}$ of N$_{2}$H$^{+}$ derived by the Nobeyama 45-m observations at each position.
As a test, we ran the program several times and compared the results.
We found that the derived column densities change within their errors, which is the major uncertainty in our discussion (Section \ref{sec:dis}).
In addition, we should note that it is common to determine the column density and $T_{\rm {ex}}$ simultaneously from more than one molecular transition lines. 
In order to check how nicely the MCMC analysis works, we independently applied the method to the HC$_{3}$N ($J=10-9$) data obtained by the Nobeyama 45 m telescope \citep{2020MNRAS.493.2395T} and the HC$_{3}$N ($J=12-11$) data convolved to the same angular resolution (21\arcsec). 
Resulting values of $T_{\rm {ex}}$ for the two transitions are consistent with each other, though $T_{\rm {ex}}$ for HC$_{3}$N ($J=10-9$) is slightly lower than that for HC$_{3}$N ($J=12-11$) by $\sim 4$ K, which may be due to the missing flux of ACA data and slight differences of upper state energies.

We also analyzed the CH$_{3}$CHO spectra with the same method (Fig.~\ref{fig:CH3CHO}).
The results of the Gaussian fitting do not seem to perfectly match the spectral shape, although we also note that the velocity resolution of the observations is 3.43 \,km\,s$^{-1}$, and better resolution is needed to reveal finer details for a better fit.

%%%%%%%%%%%%%%%%%
\begin{table*}
	%\centering
	\caption{Summary of analytical results}
	\label{tab:CASSIS1}
	\begin{tabular}{lcccc} 
		\hline
		Position & $N$ & $T_{\rm {ex}}$ & FWHM & $V_{\rm {LSR}}$ \\
		 &  (cm$^{-2}$) & (K) & (\,km\,s$^{-1}$) & (\,km\,s$^{-1}$) \\
		\hline
		\multicolumn{5}{c}{{\bf{HC$_{3}$N}}}\\
		P1 & 1.7 (0.9)$\times10^{13}$ & 14 (3) & 0.82 (0.19) & 4.77 (0.21) \\
		P2 & 1.1 (0.7)$\times10^{13}$ & 15 (3) & 0.77 (0.13) & 4.86 (0.36) \\
		P3 & 1.9 (0.7)$\times10^{13}$ & 13 (2) & 1.40 (0.19) & 4.49 (0.09) \\
		P4 & 2.0 (0.9)$\times10^{12}$ & 25 (4) & 1.03 (0.17) & 5.84 (0.27) \\
		P5 & 9.4 (3.5)$\times10^{12}$ & 14 (2) & 0.87 (0.17) & 4.99 (0.07) \\
		CH$_{3}$CHO Peak & 1.7 (0.9)$\times10^{12}$ & 17 (3) & 0.89 (0.13) & 5.68 (0.58) \\
		\multicolumn{5}{c}{{\bf{CH$_{3}$OH$^{a}$}}}\\
		P1 & 1.1 (0.3)$\times10^{14}$ & 15 (2) & 0.96 (0.13) & 5.16 (0.10) \\
		P2 & 2.4 (0.3)$\times10^{14}$ & 32 (3) & 1.87 (0.12) & 5.60 (0.06) \\
		P3 & 1.4 (0.4)$\times10^{14}$ & 15 (4) & 1.11 (0.20) & 3.98 (0.14) \\
		P4 & 2.8 (0.5)$\times10^{14}$ & 34 (5) & 2.78 (0.20) & 5.89 (0.06) \\
		P5 & 2.5 (0.9)$\times10^{14}$ & 16 (3) & 1.33 (0.16) & 5.43 (0.24) \\
		CH$_{3}$CHO Peak & 1.2 (0.7)$\times10^{14}$ & 23 (4) & 1.84 (0.33) & 5.07 (0.52) \\
		\multicolumn{5}{c}{{\bf{CH$_{3}$CN}}}\\
		P1 & 4.5 (1.0)$\times10^{11}$ & 19.9 (5.1) & 0.66 (0.08) & 4.9$^{b}$\\
		P4 & 2.9 (0.4)$\times10^{11}$ & 29.2 (1.3) & 0.85 (0.14) & 5.97 (0.05) \\
		\multicolumn{5}{c}{{\bf{CH$_{3}$CHO$^{a}$}}}\\
		CH$_{3}$CHO Peak & 6.0 (2.4)$\times10^{11}$ & 26 (9) & 2.11 (0.69) & 5.04 (0.54) \\
		\hline
		\multicolumn{4}{l}{Note: Numbers in parentheses are the standard deviation.}\\
		\multicolumn{4}{l}{$a$: Both $A$-type and $E$-type are taken into consideration,} \\
		\multicolumn{4}{l}{assuming the same column density for both types.} \\
		\multicolumn{4}{l}{$b$: The fitting was done with the fixed radial velocity.} \\
	\end{tabular}
\end{table*}
%%%%%%%%%%%%%%%%%

\section{Discussion} \label{sec:dis}

As seen in Section~\ref{sec:map}, the spatial distributions of CH$_{3}$OH are different from those of HC$_{3}$N in NGC\,2264-D.
In Section~\ref{sec:d1}, we will compare their spatial distributions in detail.
We further compare chemical compositions among the n3 field, the n5 field, and NGC\,2264-C in Section~\ref{sec:d2}.

\subsection{Comparison of Spatial Distributions of HC$_{3}$N and CH$_{3}$OH} \label{sec:d1}

%%%%%%%%
\begin{figure*}
       \begin{center}
	\includegraphics[scale=0.6]{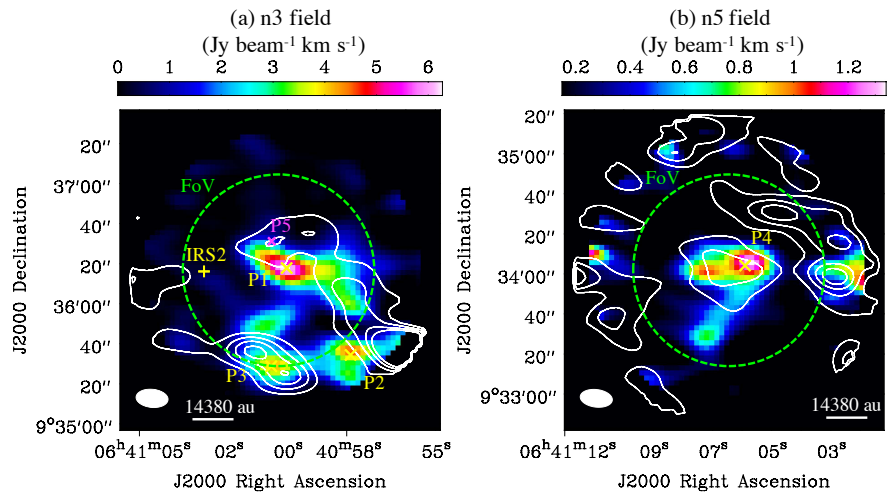}
	\end{center}
       \caption{Comparison of moment 0 maps between HC$_{3}$N (color) and CH$_{3}$OH (white contours), as in Fig.~\ref{fig:mom0}, toward (a) the n3 field and (b) the n5 field, respectively.}
    \label{fig:comparemom0}
\end{figure*}
%%%%%%%%%%

Fig.~\ref{fig:comparemom0} shows comparisons of spatial distributions between CH$_{3}$OH and HC$_{3}$N toward (a) the n3 field and (b) the n5 field, respectively.
In the n3 field, three HC$_{3}$N peaks, P1 -- P3, are not coincident with the CH$_{3}$OH emission peaks.
A bright knot and extended west-facing bow shock, labeled as MHO 1385 by \citet{2012MNRAS.423.1127B}, was identified in an H$_{2}$ map at the exact position of P1.
\citet{2012MNRAS.423.1127B} mentioned that the detection of the bright knot and bow shock indicates the presence of protostar(s).
Thus, the CH$_{3}$OH emission seems to come from shock regions induced by the jet or molecular outflows originating from protostar(s) embedded in the DMM6 core or the P1 clump.

The P2 position is located at the edge of the CH$_{3}$OH emission peak.
At this edge of the field of view, the line widths of CH$_{3}$OH are significantly broad, and the strong SO and CS lines are detected.
Thus, the CH$_{3}$OH emission is likely associated with a shock region.
Although we could not identify the origin of the HC$_{3}$N emission to be exactly at P2, it is likely that the HC$_{3}$N emission is associated with a Spitzer source (YSO) identified at 2\farcs86 from the P2 peak position \citep{2014ApJ...794..124R}.

The P3 clump corresponds to the dust clump No. 8 in \citet{2015MNRAS.453.2006B}, identified using the 850 $\micron$ continuum emission.
This clump contains a Class II protostar \citep{2015MNRAS.453.2006B}.
In addition, \citet{2018ApJ...855...45S} identified an IR cluster at the P3 clump position (4753cl1).
At the P3 position, the HC$_{3}$N peak is located between two CH$_{3}$OH peaks.
These features indicate that the warm gas around a YSO is traced by the HC$_{3}$N emission, while the molecular outflows are traced by the low upper-state-energy lines of CH$_{3}$OH.

In the n5 field, the P4 position is located at the edge of the CH$_{3}$OH emission.
The HC$_{3}$N peak (P4) is consistent with the DMM1 dust continuum core (Fig.~\ref{fig:mom0}).
As seen in the CH$_{3}$OH spectra (Fig.~\ref{fig:specCH3OH}), the CH$_{3}$OH spectra have wide line widths (Table~\ref{tab:CASSIS1}) and probably show a broadened wing emission.
The HC$_{3}$N spectra at P4 shows weak wing emission (Fig.~\ref{fig:specHC3N}) and a wider line width compared to the other positions (Table~\ref{tab:CASSIS1}).
These results indicate that HC$_{3}$N is associated with protostellar cores, and some emission may come from the molecular outflow, while the CH$_{3}$OH emission prefers to trace the molecular outflow rather than hot cores.
Protostars are likely being formed in the P4 clump, because \citet{2012MNRAS.423.1127B} identified compact knots and fingers (MHO 1388) at the P4 peak.
Furthermore, both Class 0/I and II protostars have been identified in this clump \citep{2015MNRAS.453.2006B}.
An IR cluster was also identified (4753cl2; \cite{2018ApJ...855...45S}).
Hence, this position can be considered as an active star formation site.
These protostars should produce turbulent gas, and shock regions can be invoked.

In summary, the CH$_{3}$OH emission seems to trace shock regions that the HC$_{3}$N emission does not trace.
In fact, the CH$_{3}$OH spectra at P1--P3 shows, marginally, either two velocity components or a wing emission (Fig.~\ref{fig:specCH3OH}).
On the other hand, the HC$_{3}$N emission seems to trace more quiescent regions.
The HC$_{3}$N emission associated with clumps containing low-mass YSOs has been reported in other YSOs (e.g., \cite{2021ApJ...910..141T}).
The detection of CH$_{3}$OH has been reported in the molecular outflows associated with some YSOs \citep{1998A&A...335..266B, 1998MNRAS.298..644G, 2000ApJ...545..861G}.
These results support our suggestion that the CH$_{3}$OH lines detected in the NGC\,2264-D cluster-forming clump mainly trace shock regions.
Furthermore, \citet{2019ApJ...880..138H} detected CH$_{3}$CHO in the molecular outflows of NGC 1333-IRAS2A, NGC 1333-IRAS4A, and L1157, besides CH$_{3}$OH.
The spatial coincidence of CH$_{3}$CHO and CH$_{3}$OH in the P4 clump suggests that the CH$_{3}$CHO emission also originates from the molecular outflow.

The turbulent gas seems to be produced by the star formation activity in the NGC\,2264-D cluster-forming clump.
Hence, the star formation activity, in particular the jet and the molecular outflow, significantly affects the larger-scale ($\sim 0.1$ pc) chemistry. 

\subsection{Comparison of Chemical Composition between NGC\,2264-D and NGC\,2264-C} \label{sec:d2}

%%%%%%%%%%%%%%%%%
\begin{table*}
	%\centering
	\caption{Summary of the column density ratios with respect to methanol at each position in the NGC\,2264 cluster-forming clumps}
	\label{tab:chem_com}
	\begin{tabular}{lccc} 
		\hline
		Position & HC$_{3}$N/CH$_{3}$OH & CH$_{3}$CN/CH$_{3}$OH & CH$_{3}$CHO/CH$_{3}$OH \\
		\hline
		P1 & $0.160 \pm 0.102$ & $0.004 \pm 0.002$ & ... \\
		%\multicolumn{4}{c}{} \\
		%P2 & $0.047 \pm 0.031$ & ... & ...\\
		%\multicolumn{4}{c}{} \\
		%P3 & $0.133 \pm 0.064$ & ... & ... \\
		%\multicolumn{4}{c}{} \\
		P4 & $0.007 \pm 0.003$ & $0.0010 \pm 0.0002$ & ... \\
		%\multicolumn{4}{c}{} \\
		P5 & $0.038 \pm 0.021$ & ... & ... \\
		%\multicolumn{4}{c}{} \\
		CH$_{3}$CHO Peak & $0.014 \pm 0.011$ & ... & $0.005 \pm 0.004$ \\
		NGC\,2264-C$^{a}$ & $0.089 \pm 0.019$ & $0.008 \pm 0.002$ & $0.043 \pm 0.007$ \\
		\hline
		\multicolumn{4}{l}{Note: The errors are calculated from the standard deviation of the column densities.}\\
		\multicolumn{4}{l}{$a$: The column densities are taken from \citet{2015ApJ...809..162W}; $N$(HC$_{3}$N)} \\
		\multicolumn{4}{l}{= ($1.6 \pm 0.3$)$\times 10^{14}$ cm$^{-2}$, $N$(CH$_{3}$OH) = ($1.8 \pm 0.2$)$\times 10^{15}$ cm$^{-2}$, } \\
		\multicolumn{4}{l}{$N$(CH$_{3}$CN) = ($1.5 \pm 0.4$)$\times 10^{13}$ cm$^{-2}$, and $N$(CH$_{3}$CHO) = ($7.7 \pm 0.8$)$\times 10^{13}$} \\ 
		\multicolumn{4}{l}{cm$^{-2}$, respectively.} \\
	\end{tabular}
\end{table*}
%%%%%%%%%%%%%%%%%

%%%%%%%%
\begin{figure}
       \begin{center}
	\includegraphics[bb = 0 10 283 570, scale=0.6]{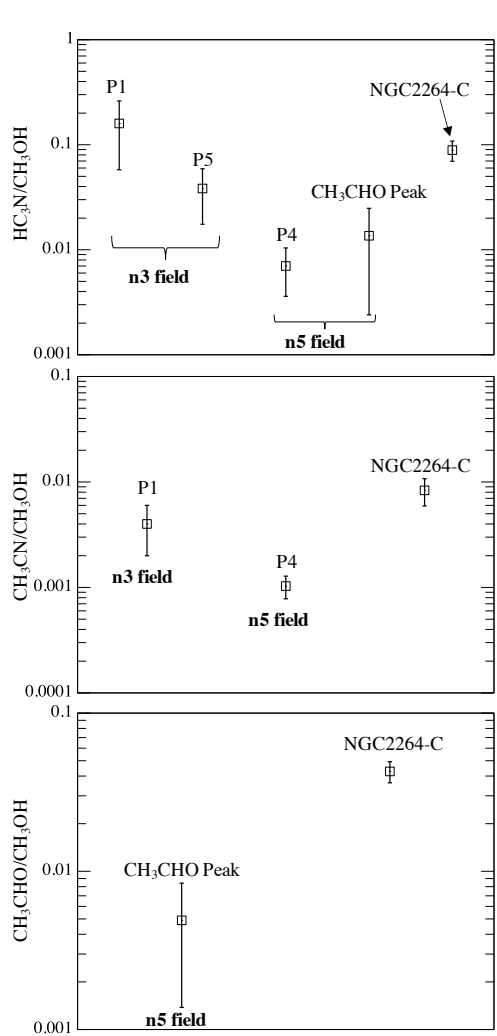}
	\end{center}
       \caption{Comparisons of chemical composition in the NGC\,2264 cluster-forming clumps. Top, middle, and bottom panels show the HC$_{3}$N/CH$_{3}$OH ratio, the CH$_{3}$CN/CH$_{3}$OH ratio, and the CH$_{3}$CHO/CH$_{3}$OH ratio, respectively.} 
    \label{fig:compare}
\end{figure}
%%%%%%%%%%

%%%%%%%%
\begin{figure}
       \begin{center}
	\includegraphics[scale=0.7]{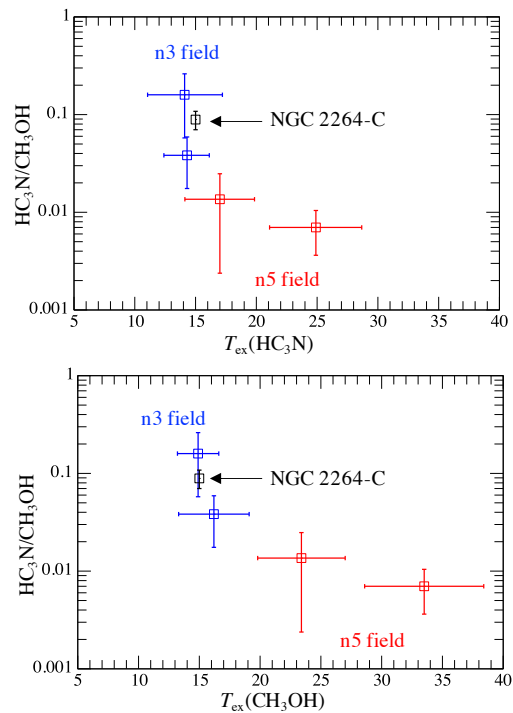}
	\end{center}
       \caption{Relationship between the excitation temperature of HC$_{3}$N and the HC$_{3}$N/CH$_{3}$OH column density ratio (upper panel). The lower panel shows the same but the excitation temperature of CH$_{3}$OH is used. The blue, red and black points indicate the n3 field, the n5 field, and CMM3 in NGC\,2264-C, respectively.}
    \label{fig:tex}
\end{figure}
%%%%%%%%%%

We derived the abundances of HC$_{3}$N, CH$_{3}$CN, and CH$_{3}$CHO with respect to CH$_{3}$OH in NGC\,2264-D, as summarized in Table~\ref{tab:chem_com}. 
We also derived these values toward CMM3 in NGC\,2264-C using observational results of \citet{2015ApJ...809..162W}.
The beam sizes of their observations were 15\arcsec--22\arcsec, which are comparable with the scale being discussed in this paper.
We adopted the CH$_{3}$OH column density of ``cold component'', ($1.8 \pm 0.2$)$\times 10^{15}$ cm$^{-2}$, because its excitation temperature of $\sim 25$ K is closer to our results (Table \ref{tab:CASSIS1}).
Even if we adopt the total CH$_{3}$OH column density of ($2.1 \pm 0.2$)$\times 10^{15}$ cm$^{-2}$ \citep{2015ApJ...809..162W}, our results and discussions do not change significantly.  
Missing flux may be present in our ACA data.
Since we do not have single-dish data of the same transition except for CH$_{3}$OH, we cannot derive missing flux for each line.
However, we can compare chemical compositions among different sources using the column density ratios, which will reduce the effect of the missing flux.

Fig.~\ref{fig:compare} shows comparisons of the abundances of three molecules with respect to CH$_{3}$OH in the NGC\,2264 cluster-forming clumps. 
The top panel shows the HC$_{3}$N case.
As mentioned in Section \ref{sec:int}, the HC$_{3}$N/CH$_{3}$OH abundance ratio is a tracer to investigate evolutionary stages of YSOs and environment.
The HC$_{3}$N/CH$_{3}$OH ratios in the n3 field have larger values than those in the n5 field systematically.
The HC$_{3}$N abundance is likely less sensitive to protostellar evolution compared to CH$_{3}$OH, because HC$_{3}$N is continually formed from CH$_{4}$ and/or C$_{2}$H$_{2}$, which are sublimated from dust grains, in warm or hot regions (25 K $<T<$ 100 K) (e.g., \cite{2019ApJ...881...57T}).
Thus, these results imply that CH$_{3}$OH is efficiently sublimated from dust grains in the n5 field.

In order to confirm this hypothesis, we investigated the relationship between the excitation temperatures of HC$_{3}$N and CH$_{3}$OH, $T_{\rm {ex}}$(HC$_{3}$N) and $T_{\rm {ex}}$(CH$_{3}$OH), and the HC$_{3}$N/CH$_{3}$OH ratio, as shown in Fig.~\ref{fig:tex}.
Since there are not enough data points, we could not conduct statistical tests.
However, both panels of Fig.~\ref{fig:tex} show a similar tendency; the HC$_{3}$N/CH$_{3}$OH ratio decreases as their excitation temperatures increase.
Even if there are uncertainties in deriving the column densities (Section \ref{sec:ana}) and the missing flux, this main conclusion does not change, because differences in the HC$_{3}$N/CH$_{3}$OH abundance ratios between n3 and n5 are around one order of magnitude.

The anti-correlations between the HC$_{3}$N/CH$_{3}$OH ratio and $T_{\rm {ex}}$ indicate that CH$_{3}$OH can be more efficiently sublimated from dust grains in warmer clumps.
These results suggest not only that thermal sublimation occurs by heating from protostars, but also by heating of dust grains in turbulent regions, as well as by the effects of photons (i.e., the photon heating effect causing non-thermal desorption) \citep{2021ApJ...910..141T}.
Overall, the n5 field is a more active star formation site than the n3 field in terms of chemical compositions.

The middle panel of Fig.~\ref{fig:compare} shows a comparison of the CH$_{3}$CN/CH$_{3}$OH ratio at P1, P4, and CMM3 in NGC\,2264-C.
This ratio at the P4 position is lower than those at the others, and the ratio at P1 is similar to that at CMM 3 in NGC\,2264-C.
This tendency is the same as the HC$_{3}$N case.
\citet{2009A&A...507L..25C} derived the CH$_{3}$CN/CH$_{3}$OH column density ratio in the L1157-B1 clumps in the molecular outflow.
The derived values in L1157 were (0.2--1.3)$\times 10^{-3}$, which is consistent with the value at P4 in the n5 field.

The bottom panel (Fig.~\ref{fig:compare}) shows a comparison of the CH$_{3}$CHO/CH$_{3}$OH ratio at the CH$_{3}$CHO Peak and CMM3 in NGC\,2264-C.
Since we detected only two CH$_{3}$CHO lines, there is a large uncertainty in the CH$_{3}$CHO abundance.
The CH$_{3}$CHO/CH$_{3}$OH ratio at the CH$_{3}$CHO peak seems to be lower than that at CMM3 in NGC\,2264-C, but similar to the values derived in molecular outflows ((0.8--6.7)$\times 10^{-3}$; \cite{2019ApJ...880..138H}). 
The results of the CH$_{3}$CN/CH$_{3}$OH ratio at P4 and the CH$_{3}$CHO/CH$_{3}$OH ratio at the CH$_{3}$CHO Peak support the view that this region is a shock region induced by the molecular outflow.

From the point of view of chemical compositions, the P4 clump is a more active star forming region than the others in NGC\,2264-D. 
\citet{2020MNRAS.493.2395T} concluded that the n5 position is the most evolved in the NGC\,2264-D cluster-forming clump by the combination of evolution along the filamentary structure and evolution of each clump.
Using these ACA data, we found that the degree of star formation activities and the nature of environments are also different among clumps in the NGC\,2264-D region in terms of chemical composition.
In fact, the most active star forming clump, the P4 clump, contains Class 0/I and II sources, while there are only Class 0/I sources in the P1 and P2 clumps, and a Class II source in the P3 clump, respectively \citep{2015MNRAS.453.2006B}.
The mixing of a variety of classes of YSOs seems to indicate the second round of star formation in the P4 clump.
These results also support the view that the P4 clump in the n5 field is the most active ongoing star formation site.

Chemical compositions in cluster-forming clumps change from site to site.
These results indicate that the stellar feedback or environments affect the chemical composition of the large clump scale.
Although the n3 field is farther from NGC\,2264-C compared to the n5 field, the chemical characteristics of the n3 field may be similar to those of CMM3 in NGC\,2264-C. 
The current data set includes a limited number of molecules, and future line survey observations with higher angular resolution are needed to confirm similarities/differences of chemical characteristics, and to disentangle the complicated effects from each protostar.

\section{Conclusions} \label{sec:con}

We present the ALMA ACA Cycle 7 data in Band 3 toward the n3 and n5 fields in the NGC\,2264-D cluster-forming region.
The HC$_{3}$N, CH$_{3}$OH, and CH$_{3}$CN lines were detected.
In addition, the CS, SO, and CH$_{3}$CHO emission lines were detected in a spectral window with the widest frequency coverage for the continuum observations.

We identified three and one HC$_{3}$N emission peaks in the n3 and n5 fields, respectively.
The strongest dust continuum core in the n3 field, DMM6, is not associated with the HC$_{3}$N peak, while the P3 clump is consistent with the DMM4 continuum core.
On the other hand, the spatial distributions of the observed CH$_{3}$OH lines are similar to those of CS and SO.
The P4 clump identified by the HC$_{3}$N emission in the n5 field is consistent with the DMM1 continuum core. 
The CH$_{3}$OH emission is also associated with the P4 clump, and its spectra show broad line widths and a wing emission, suggestive of a molecular outflow.
In general, turbulent gas induced by star formation activities produces shock regions in NGC\,2264-D.
The sites of protostars are traced by the dust continuum and HC$_{3}$N emission peaks.

We derived the HC$_{3}$N, CH$_{3}$CN, and CH$_{3}$CHO abundances with respect to CH$_{3}$OH.
In general, chemical compositions in the n3 field are more similar to those at CMM3 in NGC\,2264-C compared to the n5 field, despite the fact that the n3 field is farther from NGC\,2264-C.
We find anti-correlations between the HC$_{3}$N/CH$_{3}$OH abundance ratio and the excitation temperatures of HC$_{3}$N and CH$_{3}$OH.
The lower HC$_{3}$N/CH$_{3}$OH ratio at the P4 clump in the n5 field seems to suggest that CH$_{3}$OH is efficiently sublimated from dust grains by thermal or non-thermal heating.
These results imply that the P4 clump is a more active star formation site than other clumps in the n3 field.
This suggestion is supported by the fact that the P4 clump includes both Class 0/I and II sources, and the second round of star formation may have proceeded.

The clump scale chemistry seems to be significantly affected by star formation activities.
Our results show that the clump-scale chemical composition is a key for understanding the evolution and environments of clumps.

\begin{ack}
This paper makes use of the following ALMA data: ADS/JAO.ALMA\#2019.2.00030.S.
ALMA is a partnership of ESO (representing its member states), NSF (USA) and NINS (Japan), together with NRC (Canada), MOST and ASIAA (Taiwan), and KASI (Republic of Korea), in cooperation with the Republic of Chile. 
The Joint ALMA Observatory is operated by ESO, AUI/NRAO and NAOJ. 
Based on analyses carried out with the CASSIS software and JPL spectroscopic database. 
CASSIS has been developed by IRAP-UPS/CNRS (http://cassis.irap.omp.eu).
This work was supported by JSPS KAKENHI grant No. JP20K14523. 
This work was supported in part by Japan Foundation for Promotion of Astronomy.
E.H. thanks the National Science Foundation for support through grant AST-1906489.
We thank the anonymous referee who gave us valuable comments, which helped us improve the quality of the paper.
\end{ack}

\appendix 
\section*{Parameter ranges used in the MCMC method}

Table~\ref{tab:a1} summarizes the parameter ranges for the spectral analyses (Section \ref{sec:ana}).

%%%%%%%%%%%%%%%%%
\begin{table*}
	%\centering
	\caption{Summary of parameter ranges used in the MCMC method}
	\label{tab:a1}
	\begin{tabular}{lccc} 
		\hline
		Species & $N$  & FWHM & $V_{\rm {LSR}}$ \\
		             & (cm$^{-2}$) & (\,km\,s$^{-1}$) &  (\,km\,s$^{-1}$) \\
		\hline
		HC$_{3}$N & $5\times10^{11}-5\times10^{15}$ & $0.5-2.5$ & $4.5-5.0$ \\
		CH$_{3}$OH & $1\times10^{11}-1\times10^{16}$ & $0.5-3.0$ & $3.5-6.0$\\
		CH$_{3}$CN & $1\times10^{11}-1\times10^{14}$ & $0.5-2.5$ & $5.7-6.5$ \\
		CH$_{3}$CHO & $1\times10^{11}-1\times10^{13}$ & $1.0-4.0$ & $4.0-6.0$ \\
		\hline
		\multicolumn{4}{l}{Note: The excitation temperature in the range}\\
		\multicolumn{4}{l}{10--40 K is applied for all of the case.}\\
	\end{tabular}
\end{table*}
%%%%%%%%%%%%%%%%%

%%%

\end{document}